\documentclass[11pt]{article}
\usepackage[latin9]{inputenc}
\usepackage{textcomp}
\usepackage{amsmath}
\usepackage{amssymb}
\usepackage{graphicx}
\usepackage{esint}
\usepackage[unicode=true,
 bookmarks=false,
 breaklinks=false,pdfborder={0 0 1},backref=section,colorlinks=false]
 {hyperref}
\usepackage{breakurl}

\makeatletter

\DeclareRobustCommand{\greektext}{%
  \fontencoding{LGR}\selectfont\def\encodingdefault{LGR}}
\DeclareRobustCommand{\textgreek}[1]{\leavevmode{\greektext #1}}
\DeclareFontEncoding{LGR}{}{}
\DeclareTextSymbol{\~}{LGR}{126}
\providecommand{\tabularnewline}{\\}

\@ifundefined{date}{}{\date{}}

\usepackage{esint}
\usepackage{latexsym}
\usepackage{graphicx}\usepackage{bm}\usepackage{longtable}

\usepackage{xcolor}

\@addtoreset{equation}{section}

\makeatother

\begin{document}

\title{Holographic heavy-baryons in the Witten-Sakai-Sugimoto model with
the D0-D4 background}

\maketitle
\begin{center}
\footnote{Email: siwenli@fudan.edu.cn}Si-wen Li\emph{$^{\dagger}$}
\par\end{center}

\begin{center}
\emph{$^{\dagger}$Department of Physics,}\\
\emph{ Center for Field Theory and Particle Physics, }\\
\emph{ Fudan University, }\\
\emph{Shanghai 200433, China}
\par\end{center}

\vspace{10mm}

\begin{abstract}
We extend the holographic analysis of light-baryon spectrum in \cite{key-50}
to the case involving the heavy flavors. With the construction of
the Witten-Sakai-Sugimoto model in the D0-D4 background, we use the
mechanism proposed in \cite{key-59,key-60,key-61} by including two
light and one heavy flavor branes, to describe the heavy-light baryons
as heavy mesons bound to a flavor instanton. The background geometry
of this model corresponds to an excited state in the dual field theory
with nonzero glue condensate $\left\langle \mathrm{Tr}\mathcal{F}\wedge\mathcal{F}\right\rangle =8\pi^{2}N_{c}\tilde{\kappa}$,
or equivalently a $\theta$ angle, which is proportional to the number
density of the D0-brane charge. At strongly coupled limit, this model
shows that the heavy meson is always bound in the form of the zero
mode of the flavor instanton in the fundamental representation. We
systematically study the quantization for the effective Lagrangian
of heavy-light baryons by employing the soliton picture, and derive
the mass spectrum of heavy-light baryons in the situation with single-
and double-heavy baryon. We find the difference in the mass spectrum
becomes smaller if the density of D0-brane charge increases and the
constraint of stable states of the heavy-light baryons is $1<b<3$.
It indicates that baryon can not stably exist for sufficiently large
density of D0 charge which is in agreement with the conclusions in
the previous study of this model.
\end{abstract}
\newpage{}

\section{Introduction}

The heavy quark (c, b, t) is characterized by the heavy-quark symmetry
in QCD \cite{key-1} while the light quark (u, d, s) is dominated
by the spontaneous breaking of chiral symmetry. As measured by \cite{key-2,key-3},
the chiral doubling in heavy-light mesons \cite{key-4,key-5,key-6,key-7}
combine these symmetries. Recently, flurry of experiments report that
many multiquark exotics are incommensurate with quarkonia \cite{key-8,key-9,key-10,key-11,key-12,key-13,key-14}.
Accordingly, some new phenomena involving heavy-light multiquark states
are strongly supported by these experimental results. On the other
hand, the spontaneous parity violation in QCD has also been discussed
with the running of the RHIC in many researches \cite{key-15,key-16,key-17,key-18,key-19}.
It is well known that the $P$ or $CP$ violation could be usually
described by an nonzero $\theta$ angle in the action of such theories.
So when deconfinement happens in QCD, a metastable state with nonzero
$\theta$ angle may probably be produced in the hot and dense situation
in RHIC, then the bubble forms with odd $P$ or $CP$ parity would
soon decay into the true vacuum. For example, the chiral magnetic
effect (CME) was proposed as a test of such phenomena \cite{key-20,key-21,key-22}.
Thus it is theoretically interesting to study the $\theta$ dependence
of some observables in QCD or in the gauge theory e.g. $\theta$ dependence
of the spectrum of the glueball \cite{key-23}, the phase diagram
\cite{key-24,key-25} and the $\theta$ dependence in the large $N_{c}$
limit \cite{key-26} (one can also review the details of the $\theta$
dependence in \cite{key-27}). 

The holographic construction by the gauge/gravity duality offers a
framework to investigate the aspects of the strongly coupled quantum
field theory \cite{key-28,key-29} since QCD at low energy scale is
non-perturbative. Using the D4-brane construction in Witten's \cite{key-30},
a concrete model was proposed by Sakai and Sugimoto \cite{key-31,key-32}
which almost contains all necessary ingredients of QCD e.g. baryons
\cite{key-33,key-34}, quark matter, chiral/deconfinement phase transitions
\cite{key-35,key-36,key-37,key-38}, glueball spectrum and interaction
\cite{key-39,key-40,key-41,key-42,key-43}. Specifically, flavors
are introduced into this model by a stack of $N_{f}$ pairs of suitable
$\mathrm{D}8/\overline{\mathrm{D}8}$-branes as probes embedded in
the $N_{c}$ D4-branes background. The chiral quarks are in the fundamental
representation of the color and flavor group which come from the massless
spectrum of the open strings stretched between the color and flavor
($\mathrm{D}8/\overline{\mathrm{D}8}$) branes. Since the flavor branes
are connected, it provides a geometrical description for the spontaneous
breaking of the chiral symmetry. Baryon in this model is the D4-branes\footnote{We will use ``$\mathrm{D}4^{\prime}$-brane'' to denote the baryon
vertex in order to distinguish the $N_{c}$ D4-branes in the following
sections.} warped on $S^{4}$ which is named as ``baryon vertex'' and it has
been recognized as the instanton configuration of the gauge field
on the worldvolume of the flavor branes \cite{key-44}. In particular,
the $\theta$ angle in the dual theory is realized as the instantonic
D-brane (D-instanton) holographically in the construction of the string
theory \cite{key-45}. Hence adding the D-instanton (D0-branes) to
the background of Witten-Sakai-Sugimoto model involves the $\theta$
dependence in the dual field theory \cite{key-46,key-47}. With the
systematical study of the Witten-Sakai-Sugimoto model in the D0-D4
brane background (i.e. D0-D4/D8 brane system) in \cite{key-48,key-49},
it provides a way to investigate the $\theta$ dependence of QCD or
Yang-Mills theory holographically \cite{key-50,key-51,key-52,key-53}\footnote{See also \cite{key-54,key-55} for a similar approach or the applications
in \cite{key-56,key-57} in hydrodynamics with the model in \cite{key-48,key-49}
while the D0-brane is not D-instanton. }.

\begin{figure}
\begin{centering}
\includegraphics[scale=0.5]{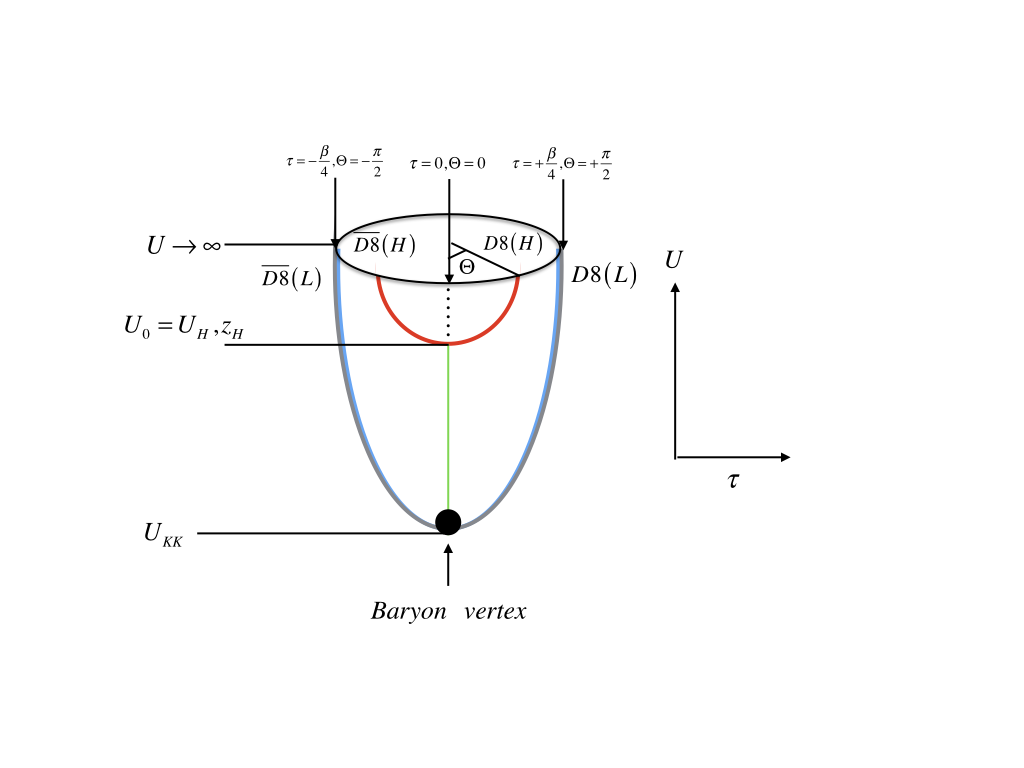}
\par\end{centering}

\caption{\label{fig:1}Various brane configuration in the $\tau-U$ plane where
$\tau$ is compactified on $S^{1}$. The bubble background (cigar)
is produced by $N_{c}$ D4-branes with $N_{0}$ smeared D0-branes.
The $N_{f}=2$ light flavor $\mathrm{D}8/\overline{\mathrm{D}8}$-branes
(L) living at antipodal position of the cigar are represented by the
blue line. A pair of heavy flavor $\mathrm{D}8/\overline{\mathrm{D}8}$-brane
(H) is separated from the light flavor branes which is represented
by the red line. The massive state is produced by the string stretched
between the light and heavy flavor branes (HL-string) which is represented
by the green line in this figure.}
\end{figure}

Since the baryon spectrum with light flavors can be reviewed in \cite{key-50,key-51,key-54,key-55}
by using the Witten-Sakai-Sugimoto model with the D0-D4 brane background,
naturally the purpose of this paper is to extend the analysis to involve
the heavy flavors. As mentioned, the fundamental quarks in this model
are represented as the fermion states created by the open strings
stretched between the $N_{c}$ D4-branes and $N_{f}$ coincident $\mathrm{D}8/\overline{\mathrm{D}8}$-branes
without length which are therefore massless states \cite{key-31,key-58}.
So the fundamental quarks in this model can be named as ``light quarks''
and the $N_{f}$ coincident flavor branes are certainly named as ``light
flavor branes''. In order to address the chiral and heavy-quark symmetries
by the holographic duality, we follow the mechanism proposed in \cite{key-59,key-60,key-61},
that is to consider one pair of flavor brane as probe (named as ``heavy
flavor brane'') which is separated from the other $N_{f}$ coincident
flavor branes as shown in Figure \ref{fig:1}. The string stretched
between the heavy and light flavor branes (named as ``HL-string'')
produces massive multiplets. Hence the heavy-light mesons correspond
to the low energy modes of these strings which can be approximated
by the local vector fields in bi-fundamental representation in the
vicinity of the light flavor branes. With the nonzero vacuum expectation
value (vev), the heavy-light fields are massive and their mass comes
from the moduli span by the dilaton fields in the action. This setup
allows to describe the radial spectra of heavy-light multiplets, their
pertinent vector and axial correlations and so on. The baryons with
heavy flavor should be the form of instanton configurations in the
worldvolume theory of $\mathrm{D}8/\overline{\mathrm{D}8}$-branes
bound to heavy-light vector mesons. And this method will also develop
the bound state approach in the context of the Skyrme model (e.g.
\cite{key-62}) holographically by involving the $\theta$ dependence. 

The outline of this paper is as follows. In section 2, we briefly
review the D0-D4 background and its dual field theory. In section
3, we outline the geometrical setup and derive the heavy-light effective
action through the Dirac-Born-Infield (DBI) and the Chern-Simons (CS)
actions of the D-branes. It shows the heavy-meson interactions to
the instanton on the flavor branes. In section 4, we show the effective
action in the double limit and a spin-1 vector meson is binding to
the bulk instanton transmuting to spin-1/2. In section 5, we employ
the quantization and show how to involve the heavy flavors in the
spectrum additional to the previous works \cite{key-50,key-51,key-54,key-55}
with a finite D0-brane charge or $\theta$ angle. The derivation of
baryon spectra with single and double-heavy quarks are explicitly
given in this section. Section 6 is the summary. In the appendix,
we briefly summarize the essential steps for the quantization of the
light meson moduli without the heavy flavor branes which has already
been studied in \cite{key-50}.

\section{The D0-D4 background and the dual field theory}

In this section, we will review the D0-D4 background and its dual
field theory by following \cite{key-48}. In Einstein frame, the solution
of D4-brane with smeared D0-brane in the IIA supergravity is given
as,

\begin{equation}
ds^{2}=H_{4}^{-\frac{3}{8}}\left[-H_{0}^{\frac{7}{8}}f\left(U\right)d\tau^{2}+H_{0}^{\frac{1}{8}}\delta_{\mu\nu}dx^{\mu}dx^{\nu}\right]+H_{4}^{\frac{5}{8}}H_{0}^{\frac{1}{8}}\left[\frac{dU^{2}}{f\left(U\right)}+U^{2}d\Omega_{4}^{2}\right].
\end{equation}
The direction $\tau$ is compactified on a cycle with the period $\beta$.
The dilaton, Ramond-Ramond 2- and 4-form are given as,

\begin{equation}
e^{-\Phi}=g_{s}^{-1}\left(\frac{H_{4}}{H_{0}^{3}}\right)^{\frac{1}{4}},\ \ f_{2}=\frac{\left(2\pi l_{s}\right)^{7}g_{s}N_{0}}{\omega_{4}V_{4}}\frac{1}{U^{4}H_{0}^{2}}dU\wedge d\tau,\ \ f_{4}=\frac{\left(2\pi l_{s}\right)^{3}N_{c}g_{s}}{\omega_{4}}\epsilon_{4},\label{eq:2}
\end{equation}
where,

\begin{align}
H_{4}=1+\frac{U_{Q4}^{3}}{U^{3}},\ \ H_{0}=1+\frac{U_{Q0}^{3}}{U^{3}},\ \ f\left(U\right)=1-\frac{U_{KK}^{3}}{U^{3}},\nonumber \\
U_{Q0}^{3}=\frac{1}{2}\left(-U_{KK}^{3}+\sqrt{U_{KK}^{6}+\left(\left(2\pi\right)^{5}l_{s}^{7}g_{s}\tilde{\kappa}N_{c}\right)^{2}}\right),\nonumber \\
U_{Q4}^{3}=\frac{1}{2}\left(-U_{KK}^{3}+\sqrt{U_{KK}^{6}+\left(\left(2\pi\right)^{5}l_{s}^{7}g_{s}N_{c}\right)^{2}}\right).
\end{align}
We have used $d\Omega_{4}$, $\epsilon_{4}$ and $\omega_{4}=8\pi^{2}/3$
to represent the line element, the volume form and the volume of a
unit $S^{4}$. $U_{KK}$ is the horizon position of the radius coordinate
and $V_{4}$ is the volume of the D4-branes. The numbers of D4- and
D0-branes are denoted by $N_{c}$ and $N_{0}$ respectively. D0-branes
are smeared in the directions of $x^{0},x^{1},x^{2},x^{3}$. So the
number density of the D0-branes can be represented by $N_{0}/V_{4}$.
In order to take account of the backreaction from the D0-branes as
\cite{key-47}, we also require that $N_{0}$ is order of $N_{c}$.
In the large $N_{c}$ limit, $\tilde{\kappa}$ would be order of $\mathcal{O}\left(1\right)$
which is defined as $\tilde{\kappa}=N_{0}/\left(N_{c}V_{4}\right)$.

In the string frame, making double Wick rotation  and taking field
limit i.e. $\alpha^{\prime}\rightarrow0$ with fixed $U/\alpha^{\prime}$
and $U_{KK}/\alpha^{\prime}$, then we obtain the D0-D4 bubble geometry,
the metric becomes,

\begin{align}
ds^{2}= & \left(\frac{U}{R}\right)^{3/2}\left[H_{0}^{1/2}\eta_{\mu\nu}dx^{\mu}dx^{\nu}+H_{0}^{-1/2}f\left(U\right)d\tau^{2}\right]\nonumber \\
 & +H_{0}^{1/2}\left(\frac{R}{U}\right)^{3/2}\left[\frac{dU^{2}}{f\left(U\right)}+U^{2}d\Omega_{4}^{2}\right],\label{eq:4}
\end{align}
and the dilaton become,

\begin{equation}
e^{\Phi}=g_{s}\left(\frac{U}{R}\right)^{3/4}H_{0}^{3/4},\label{eq:5}
\end{equation}
where $R^{3}=\pi g_{s}l_{s}^{3}N_{c}$ is the limit of $U_{Q4}^{3}$.
Here $l_{s}$ is the length of the string and $\alpha^{\prime}=l_{s}^{2}$.
In the bubble geometry (\ref{eq:4}), the spacetime ends at $U=U_{KK}$.
In order to avoid the conical singularity at $U_{KK}$, the period
$\beta$ of $\tau$ must satisfy,

\begin{equation}
\beta=\frac{4\pi}{3}U_{KK}^{-1/2}R^{3/2}b^{1/2},\ \ b\equiv H_{0}\left(U_{KK}\right).\label{eq:6}
\end{equation}

In the low-energy effective description, the dual theory is a five-dimensional
$U\left(N_{c}\right)$ Yang-Mills (YM) theory which lives inside the
world volume of D4-brane. Since one direction of the D4-branes is
campactified on a cycle $\tau$, the four-dimensional Yang-Mills coupling
could be obtained as studied in \cite{key-30}, which is relating
the D4-brane tension, the five-dimensional Yang-Mills coupling constant
$g_{5}$, then analyzing the relation of the five-dimensionally compactified
theory and four dimensions on the $\tau$ direction. Thus the resultantly
four-dimensional Yang-Mills coupling is,

\begin{equation}
g_{YM}^{2}=\frac{g_{5}^{2}}{\beta}=\frac{4\pi^{2}g_{s}l_{s}}{\beta},
\end{equation}
then $b$ and $R^{3}$ can be evaluated as,

\begin{equation}
b=\frac{1}{2}\left[1+\left(1+C\beta^{2}\right)^{1/2}\right],\ \ C\equiv\left(2\pi l_{s}^{2}\right)^{6}\lambda^{2}\tilde{\kappa}/U_{KK}^{6},\ \ R^{3}=\frac{\beta\lambda l_{s}^{2}}{4\pi},\label{eq:8}
\end{equation}
where the 't Hooft coupling $\lambda$ is defined as $\lambda=g_{YM}^{2}N_{c}$.
The Kaluza-Klein (KK) modes can be introduced by defining a mass scale
$M_{KK}=2\pi/\beta$. The fermion and scalar become massive at the
KK mass scale since the anti-periodic condition is imposed on the
fermions \cite{key-31}. Therefore, the massless modes of the open
string dominate the dynamics in the low-energy theory which is described
by a pure Yang-Mills theory. According to (\ref{eq:6}) and (\ref{eq:8}),
we have the following relations,

\begin{equation}
\beta=\frac{4\pi\lambda l_{s}^{2}}{9U_{KK}}b,\ \ \ \ M_{KK}=\frac{9U_{KK}}{2\lambda l_{s}^{2}b}.\label{eq:9}
\end{equation}
 Because $b\geq1$ and $U_{KK}\geq2\lambda l_{s}^{2}M_{KK}/9$, $\beta$
can be solved by using (\ref{eq:8}) and (\ref{eq:9}) as,

\begin{equation}
\beta=\frac{4\pi\lambda l_{s}^{2}}{9U_{KK}}\frac{1}{1-\frac{\left(2\pi l_{s}^{2}\right)^{8}}{81U_{KK}^{8}}\lambda^{4}\tilde{\kappa}^{2}},\ \ b=\frac{1}{1-\frac{\left(2\pi l_{s}^{2}\right)^{8}}{81U_{KK}^{8}}\lambda^{4}\tilde{\kappa}^{2}}.
\end{equation}

Let us consider the effective action of a D4-brane with the smeared
D0-branes in the background which takes the following form,

\begin{equation}
S_{D_{4}}=-\mu_{4}\mathrm{Tr}\int d^{4}xd\tau e^{-\phi}\sqrt{-\det\left(G+2\pi\alpha^{\prime}\mathcal{F}\right)}+\mu_{4}\int C_{5}+\frac{1}{2}\left(2\pi\alpha^{\prime}\right)^{2}\mu_{4}\int C_{1}\wedge\mathcal{F}\wedge\mathcal{F},\label{eq:11}
\end{equation}
where $\mu_{4}=\left(2\pi\right)^{-4}l_{s}^{-5}$, $\phi=\Phi-\Phi_{0}$,
$e^{\Phi_{0}}=g_{s}$ and $G$ is the induced metric on the world
volume of D4-branes. $\mathcal{F}$ is the gauge field strength on
the D4-brane. $C_{5},\ C_{1}$ is the Ramond-Ramond 5- and 1- form
respectively and their field strengths are given in (\ref{eq:2}).
The Yang-Mills action can be obtained by the leading-order expansion
respected to small $\mathcal{F}$ from the first term in (\ref{eq:11})
(i.e. the DBI action). In the bubble D0-D4 solution, we have $C_{1}\sim\theta d\tau$
in (\ref{eq:2}), thus D0-branes are actually D-instantons (as shown
in Table \ref{tab:1}) and the last term in (\ref{eq:12}) could be
integrated as,

\begin{equation}
\int_{S_{\tau}^{1}}C_{1}\sim\theta\sim\tilde{\kappa},\ \ \int_{S_{\tau}^{1}\times\mathbb{R}^{4}}C_{1}\wedge\mathcal{F}\wedge\mathcal{F}\sim\theta\int_{\mathbb{R}^{4}}\mathcal{F}\wedge\mathcal{F}.\label{eq:12}
\end{equation}
So the free parameter $\tilde{\kappa}$ (related to the $\theta$
angle in the dual field theory) has been introduced into the Witten-Sakai-Sugimoto
model by this string theory background, however this background is
not dual to the vacuum state of the gauge field theory. Similarly
as in \cite{key-45}, in the dual field theory, some excited states
with a constant homogeneous field strength background may be described
in the D0-D4 model. The expectation value of $\mathrm{Tr}\mathcal{F}\wedge\mathcal{F}$
can be evaluated as $\left\langle \mathrm{Tr}\mathcal{F}\wedge\mathcal{F}\right\rangle =8\pi^{2}N_{c}\tilde{\kappa}$
\cite{key-48,key-50}. Then the deformed relations in the presence
of D0-branes of the variables in QCD are given as follows,
\begin{equation}
R^{3}=\frac{\lambda l_{s}^{2}}{2M_{KK}},\ \ g_{s}=\frac{\lambda}{2\pi M_{KK}N_{c}l_{s}},\ \ U_{KK}=\frac{2}{9}M_{KK}\lambda l_{s}^{2}b.\label{eq:13}
\end{equation}

\section{Holographic effective action for heavy-light interaction}

\subsection{D-brane setup}

The chiral symmetry $U_{R}\left(N_{f}\right)\times U_{L}\left(N_{f}\right)$
can be introduced into the D0-D4 system by adding a stack of probe
$N_{f}$ D8-anti-D8 ($\mathrm{D}8/\overline{\mathrm{D}8}$) branes
into the background which are named as flavor branes. The spontaneously
breaking of $U_{R}\left(N_{f}\right)\times U_{L}\left(N_{f}\right)$
symmetry to $U_{V}\left(N_{f}\right)$ in the dual field theory can
be geometrically understood by the separately faraway $\mathrm{D}8/\overline{\mathrm{D}8}$-branes
combining near the bottom of the bubble at $U=U_{KK}$ (as shown by
the blue lines in Figure \ref{fig:1}). This can be verified by the
appearance of massless Goldstones \cite{key-63}. The brane configurations
are illustrated in Table \ref{tab:1}. 

\begin{table}
\begin{centering}
\begin{tabular}{|c|c|c|c|c|c|c|c|c|c|c|}
\hline 
 & 0 & 1 & 2 & 3 & 4$\left(\tau\right)$ & 5$\left(U\right)$ & 6 & 7 & 8 & 9\tabularnewline
\hline 
\hline 
Smeared D0-branes & = & = & = & = & - &  &  &  &  & \tabularnewline
\hline 
$N_{c}$ D4-branes & - & - & - & - & - &  &  &  &  & \tabularnewline
\hline 
$N_{f}$ $\mathrm{D}8/\overline{\mathrm{D}8}$-branes & - & - & - & - &  & - & - & - & - & -\tabularnewline
\hline 
Baryon vertex $\mathrm{D4}^{\prime}$-branes & - &  &  &  &  &  & - & - & - & -\tabularnewline
\hline 
\end{tabular}
\par\end{centering}

\caption{\label{tab:1}The brane configurations: \textquotedblleft =\textquotedblright{}
denotes the smeared directions, \textquotedblleft -\textquotedblright{}
denotes the world volume directions.}

\end{table}

The induced metric on the probe $\mathrm{D}8/\overline{\mathrm{D}8}$-branes
is,

\begin{align}
ds_{\mathrm{D}8/\overline{\mathrm{D}8}}^{2}= & \left(\frac{U}{R}\right)^{3/2}H_{0}^{-1/2}\left[f\left(U\right)+\left(\frac{R}{U}\right)^{3}\frac{H_{0}}{f\left(U\right)}U^{\prime2}\right]d\tau^{2}\nonumber \\
 & +\left(\frac{U}{R}\right)^{3/2}H_{0}^{1/2}\eta_{\mu\nu}dx^{\mu}dx^{\nu}+H_{0}^{1/2}\left(\frac{R}{U}\right)^{3/2}U^{2}d\Omega_{4}^{2}.
\end{align}
where $U^{\prime}$ is the derivative with respect to $\tau$. The
action of the $\mathrm{D}8/\overline{\mathrm{D}8}$-branes can be
obtained as,

\begin{equation}
S_{\mathrm{D}8/\overline{\mathrm{D}8}}\propto\int d^{4}xdUH_{0}\left(U\right)U^{4}\left[f\left(U\right)+\left(\frac{R}{U}\right)^{3}\frac{H_{0}}{f\left(U\right)}U^{\prime2}\right]^{1/2},\label{eq:15}
\end{equation}
then the equation of motion for $U\left(\tau\right)$ can be derived
as,

\begin{equation}
\frac{d}{d\tau}\left(\frac{H_{0}\left(U\right)U^{4}f\left(U\right)}{\left[f\left(U\right)+\left(\frac{R}{U}\right)^{3}\frac{H_{0}}{f\left(U\right)}U^{\prime2}\right]^{1/2}}\right)=0,\label{eq:16}
\end{equation}
which can be interpreted as the the conservation of the energy. With
the initial conditions $U\left(\tau=0\right)=U_{0}$ and $U^{\prime}\left(\tau=0\right)=0$,
the generic formula of the embedding function $\tau\left(U\right)$
can be solved as,

\begin{equation}
\tau\left(U\right)=E\left(U_{0}\right)\int_{U_{0}}^{U}dU\frac{H_{0}^{1/2}\left(U\right)\left(\frac{R}{U}\right)^{3/2}}{f\left(U\right)\left[H_{0}^{2}\left(U\right)U^{8}f\left(U\right)-E^{2}\left(U_{0}\right)\right]^{1/2}},\label{eq:17}
\end{equation}
where $E\left(U_{0}\right)=H_{0}\left(U_{0}\right)U_{0}^{4}f^{1/2}\left(U_{0}\right)$
and $U_{0}$ denotes the connected position of the $\mathrm{D}8/\overline{\mathrm{D}8}$-branes.
Following \cite{key-31,key-48}, we introduce the new coordinates
$\left(r,\Theta\right)$ and $\left(y,z\right)$ which satisfy,

\begin{align}
y=r\cos\Theta, & \ \ z=r\sin\Theta,\nonumber \\
U^{3}=U_{KK}^{3}+U_{KK}r^{2}, & \ \ \Theta=\frac{2\pi}{\beta}\tau=\frac{3}{2}\frac{U_{KK}^{1/2}}{R^{3/2}H_{0}^{1/2}\left(U_{KK}\right)}.
\end{align}
In this manuscript, we will consider the following configuration for
the various flavor branes: the light flavor branes live at the antipodal
position as in \cite{key-31,key-48,key-49} which means they ($\mathrm{D}8/\overline{\mathrm{D}8}$-branes)
are embedded at $\Theta=\pm\frac{1}{2}\pi$ respectively i.e. $y=0$.
The embedding function of the light flavor branes is $\tau_{L}\left(U\right)=\frac{1}{4}\beta$
so that we have $U^{3}=U_{KK}^{3}+U_{KK}z^{2}$ on the light $\mathrm{D}8/\overline{\mathrm{D}8}$-branes\footnote{With the suitable boundary condition, $\tau_{L}\left(U\right)=\frac{1}{4}\beta$
is indeed a solution of (\ref{eq:16}) as discussed in \cite{key-31,key-48,key-49}.}. Therefore the induced metric on them becomes,

\begin{equation}
ds_{\mathrm{Light}-\mathrm{D}8/\overline{\mathrm{D}8}}^{2}=H_{0}^{1/2}\left(\frac{U}{R}\right)^{3/2}\eta_{\mu\nu}dx^{\text{\textmu}}dx^{\nu}+\frac{4}{9}\frac{U_{KK}}{U}\left(\frac{U}{R}\right)^{3/2}H_{0}^{1/2}dz^{2}+H_{0}^{1/2}\left(\frac{R}{U}\right)^{3/2}U^{2}d\Omega_{4}^{2}.\label{eq:19}
\end{equation}

For the heavy flavor branes, we have to choose another solution as
$\tau_{H}\left(U\right)$ from (\ref{eq:17}) with $U_{0}=U_{H}\neq U_{KK}$
since they must live at the non-antipodal position of the background\footnote{Since we are going to discuss the limit $U_{H},z_{H}\rightarrow\infty$
in the next section, an analytical solution for the embedding function
of the heavy flavor brane could be $\tau_{H}\left(U\right)=-\frac{2}{9}\left(\frac{R}{U}\right)^{3/2}\frac{U_{H}}{U^{3}}\ _{2}F_{1}\left(\frac{1}{2},\frac{9}{16},\frac{25}{16},\frac{U_{H}^{8}}{U^{8}}\right)+\frac{2\sqrt{\pi}}{9}\frac{R^{3/2}}{U_{H}^{1/2}}\frac{\Gamma\left(\frac{25}{16}\right)}{\Gamma\left(\frac{17}{16}\right)}$
where $\ _{2}F_{1}$ is the hypergeometric function. In this limit,
the integral region on the heavy flavor brane is $U>U_{H}\rightarrow\infty$,
so we have $f,H_{0}\sim1$ to get this solution with (\ref{eq:17}).}. So the heavy flavor brane is separated from the light flavor branes
with a finite separation at $\tau_{H}\left(U_{0}\right)=0$ as shown
in Figure \ref{fig:1}. With the approach presented in \cite{key-44,key-50,key-64},
the light baryons spectrum in the D0-D4/D8 system has been studied
in \cite{key-50,key-51,key-54,key-55}. In this paper, we extend our
previous work by following \cite{key-59,key-60,key-61} to study the
heavy-light interaction and baryon spectrum with heavy flavors in
the D0-D4/D8 system. Thus we consider $N_{f}=2$ light flavor $\mathrm{D}8/\overline{\mathrm{D}8}$-branes
(L) and one pair of heavy (H) flavor brane as probe in the bubble
D0-D4 geometry (\ref{eq:4}) that spontaneously breaks chiral symmetry.
The massive states on the light $\mathrm{D}8/\overline{\mathrm{D}8}$-branes
are produced by the heavy-light (HL) strings connecting heavy-light
branes.

\subsection{Yang-Mills and Chern-Simons action of the flavor branes}

Since the baryon vertex lives inside the light flavor branes, the
concern of this section is to study the effective dynamics of the
baryons or mesons on the light flavor branes involving the heavy-light
interaction. The lowest modes of the open string stretched between
the heavy and light branes are attached to the baryon vertex as shown
in Figure \ref{fig:1}. In our D0-D4/D8 system, these string modes
consist of longitudinal modes $\Phi_{a}$ and the transverse modes
$\Psi$ near the light brane world volume. These fields acquire a
nonzero vev at finite brane separation, which introduces the mass
to the vector field \cite{key-65}. These fields are always named
as ``bi-local'', however we will approximate them near the light
flavor branes by local vector fields hence they are described by the
standard DBI action. So this construction is distinct from the approaches
presented in \cite{key-66,key-67,key-68,key-69,key-70,key-71,key-72}.

By keeping these in mind, let us consider action of the light flavor
branes. For the D8-branes, the generic expansion of the DBI in the
leading order can be written as,

\begin{equation}
S_{\mathrm{DBI}}^{\mathrm{D8/\overline{D8}}}=-\frac{T_{8}\left(2\pi\alpha^{\prime}\right)^{2}}{4}\int d^{9}\xi\sqrt{-\det G}e^{-\Phi}\mathrm{Tr}\left\{ \mathcal{F}_{ab}\mathcal{F}^{ab}-2D_{a}\varphi^{I}D_{a}\varphi^{J}+\left[\varphi^{I},\varphi^{J}\right]^{2}\right\} .\label{eq:20}
\end{equation}
where $\varphi^{I}$ is the transverse mode of the flavor branes and
the index $a,b$ runs over the flavor brane. Notice that there is
only one transverse coordinate to the D8-brane, thus we define $\varphi^{I}\equiv\Psi$
to omit the index. The scalar field $\Psi$ is traceless and adjoint
representation additional to the adjoint gauge field $\mathcal{A}_{a}$.
Since the one pair of the heavy flavor brane is separated from the
$N_{f}=2$ light flavor branes with a string stretched between them,
in string theory the world volume field can be combined in a superconnection.
For the gauge field, we can use the following matrix-valued 1-form
as,

\begin{equation}
\boldsymbol{\mathcal{A}}_{a}=\left(\begin{array}{cc}
\mathcal{A}_{a} & \varPhi_{a}\\
-\varPhi_{a}^{\dagger} & 0
\end{array}\right),\label{eq:21}
\end{equation}
where $\boldsymbol{\mathcal{A}}_{a}$ is $\left(N_{f}+1\right)\times\left(N_{f}+1\right)$
matrix-valued while $\Psi$ and $\mathcal{A}_{a}$ are $N_{f}\times N_{f}$
valued. If all the flavor branes are coincident, the $\varPhi_{a}$
multiplet is massless, otherwise $\varPhi_{a}$ could be massive field.
The corresponding gauge field strength of (\ref{eq:21}) is,

\begin{equation}
\boldsymbol{\mathcal{F}}_{ab}=\left(\begin{array}{cc}
\mathcal{F}_{ab}-\varPhi_{[a}\varPhi_{b]}^{\dagger} & \partial_{[a}\varPhi_{b]}+\mathcal{A}_{[a}\varPhi_{b]}\\
\partial_{[a}\varPhi_{b]}^{\dagger}+\varPhi_{[a}^{\dagger}\mathcal{A}_{b]} & -\varPhi_{[a}^{\dagger}\varPhi_{b]}
\end{array}\right).\label{eq:22}
\end{equation}
Inserting the induced metric (\ref{eq:19}) with (\ref{eq:5}) into
(\ref{eq:20}), we can write the DBI action as two parts,

\begin{equation}
S_{\mathrm{DBI}}^{\mathrm{D8/\overline{D8}}}=S_{YM}^{\mathrm{D8/\overline{D8}}}+S_{\Psi}.\label{eq:23}
\end{equation}
The Yang-Mills part is calculated as,

\begin{equation}
S_{YM}=-2\tilde{T}U_{KK}^{-1}\int d^{4}xdzH_{0}^{1/2}\mathrm{Tr}\left[\frac{1}{4}\frac{R^{3}}{U}\boldsymbol{\mathcal{F}}_{\mu\nu}\boldsymbol{\mathcal{F}}^{\mu\nu}+\frac{9}{8}\frac{U^{3}}{U_{KK}}\boldsymbol{\mathcal{F}}_{\mu z}\boldsymbol{\mathcal{F}}^{\mu z}\right],\label{eq:24}
\end{equation}
where $\mu,\nu$ rum over $0,1,2,3$ and

\begin{equation}
\tilde{T}=\frac{\left(2\pi\alpha^{\prime}\right)^{2}}{3g_{s}}T_{8}\omega_{4}U_{KK}^{3/2}R^{3/2}=\frac{M_{KK}^{2}\lambda N_{c}b^{3/2}}{486\pi^{3}}.
\end{equation}
In order to work with the dimensionless variables, we introduce the
replacement $z\rightarrow zU_{KK}$, $x^{\mu}\rightarrow x^{\mu}/M_{KK}$,
$\boldsymbol{\mathcal{A}}_{z}\rightarrow\boldsymbol{\mathcal{A}}_{z}/U_{KK}$,
$\boldsymbol{\mathcal{A}}_{\mu}\rightarrow\boldsymbol{\mathcal{A}}_{\mu}M_{KK}$\footnote{Working with this replacement is equivalent to work in the unit of
$U_{KK}=M_{KK}=1$ as \cite{key-44} in this model.}, then (\ref{eq:24}) takes the following formulas,

\begin{align}
S_{YM}^{\mathrm{D8/\overline{D8}}} & =-\tilde{T}M_{KK}^{-2}\frac{9}{4b}\int d^{4}xdzH_{0}^{1/2}\left(U\right)\mathrm{Tr}\left[\frac{1}{2}\frac{U_{KK}}{U}\boldsymbol{\mathcal{F}}_{\mu\nu}^{2}+\frac{U^{3}}{U_{KK}^{3}}b\boldsymbol{\mathcal{F}}_{\mu z}^{2}\right],\nonumber \\
 & =-a\lambda N_{c}b^{1/2}\int d^{4}xdzH_{0}^{1/2}\left(U\right)\mathrm{Tr}\left[\frac{1}{2}K\left(z\right)^{-1/3}\eta^{\mu\nu}\eta^{\rho\sigma}\boldsymbol{\mathcal{F}}_{\mu\rho}\boldsymbol{\mathcal{F}}_{\nu\sigma}+K\left(z\right)b\eta^{\mu\nu}\boldsymbol{\mathcal{F}}_{\mu z}\boldsymbol{\mathcal{F}}_{\nu z}\right],\label{eq:26}
\end{align}
where $a=\frac{1}{216\pi^{3}}$ and $K\left(z\right)=1+z^{2}$. To
see the dependence of $\lambda$, it would be convenient to employ
the rescaling used in \cite{key-44}, which is

\begin{equation}
\left(x^{0},x^{M}\right)\rightarrow\left(x^{0},\lambda^{-1/2}x^{M}\right),\ \left(\mathcal{A}_{0},\Phi_{0}\right)\rightarrow\left(\mathcal{A}_{0},\Phi_{0}\right),\ \left(\mathcal{A}_{M},\varPhi_{M}\right)\rightarrow\left(\lambda^{1/2}\mathcal{A}_{M},\lambda^{1/2}\varPhi_{M}\right),\label{eq:27}
\end{equation}
where $M,N$ run over $1,2,3,z$ and $i,j=1,2,3$. Using (\ref{eq:27})
in the large $\lambda$ limit, (\ref{eq:26}) becomes,

\begin{align}
\mathcal{L}_{YM}^{\mathrm{D8/\overline{D8}}} & =-aN_{c}b^{3/2}\mathrm{Tr}\left[\frac{\lambda}{2}\boldsymbol{\mathcal{F}}_{MN}^{2}-bz^{2}\left(\frac{5}{12}-\frac{1}{4b}\right)\boldsymbol{\mathcal{F}}_{ij}^{2}+\frac{bz^{2}}{2}\left(1+\frac{1}{b}\right)\boldsymbol{\mathcal{F}}_{iz}^{2}-\boldsymbol{\mathcal{F}}_{0M}^{2}\right]\nonumber \\
 & \equiv aN_{c}b^{3/2}\mathcal{L}_{YM}^{L}+aN_{c}b^{3/2}\lambda\mathcal{L}_{0}^{H}+aN_{c}b^{3/2}\mathcal{L}_{1}^{H}+\mathcal{O}\left(\lambda^{-1}\right),\label{eq:28}
\end{align}
where $\mathcal{L}_{YM}^{L}$ represents the Lagrangian for the light
hadrons which has been derived in \cite{key-50} and the explict form
of $\mathcal{L}_{YM}^{L}$ can also be found in (\ref{eq:77}) (\ref{eq:78})
in the appendix. Substituting (\ref{eq:28}) for (\ref{eq:22}), we
obtain

\begin{align}
\mathcal{L}_{0}^{H}= & -\left(D_{M}\varPhi_{N}^{\dagger}-D_{N}\varPhi_{M}^{\dagger}\right)\left(D_{M}\varPhi_{N}-D_{N}\varPhi_{M}\right)+2\varPhi_{M}^{\dagger}\mathcal{F}^{MN}\varPhi_{N},\nonumber \\
\mathcal{L}_{1}^{H}= & 2\left(D_{0}\varPhi_{M}^{\dagger}-D_{M}\varPhi_{0}^{\dagger}\right)\left(D_{0}\varPhi_{M}-D_{M}\varPhi_{0}\right)-2\varPhi_{0}^{\dagger}\mathcal{F}^{0M}\varPhi_{M}-2\varPhi_{0}^{\dagger}\mathcal{F}^{0M}\varPhi_{M}+\widetilde{\mathcal{L}}_{1}^{H},\label{eq:29}
\end{align}
where $D_{M}\varPhi_{N}=\partial_{M}\varPhi_{N}+\mathcal{A}_{[M}\varPhi_{N]}$
and

\begin{align}
\widetilde{\mathcal{L}}_{1}^{H}= & bz^{2}\left(\frac{5}{6}-\frac{1}{2b}\right)\left(D_{i}\varPhi_{j}-D_{j}\varPhi_{i}\right)^{\dagger}\left(D_{i}\varPhi_{j}-D_{j}\varPhi_{i}\right)\nonumber \\
 & -bz^{2}\left(1+\frac{1}{b}\right)\left(D_{i}\varPhi_{z}-D_{z}\varPhi_{i}\right)^{\dagger}\left(D_{i}\varPhi_{z}-D_{z}\varPhi_{i}\right)\nonumber \\
 & -bz^{2}\left(\frac{5}{3}-\frac{1}{b}\right)\varPhi_{i}^{\dagger}\mathcal{F}^{ij}\varPhi_{j}+bz^{2}\left(1+\frac{1}{b}\right)\left(\varPhi_{z}^{\dagger}\mathcal{F}^{zi}\varPhi_{i}+c.c.\right).\label{eq:30}
\end{align}

The action $S_{\Psi}$ in (\ref{eq:23}) is collected as,

\begin{align}
S_{\Psi} & =-\frac{T_{8}\left(2\pi\alpha^{\prime}\right)^{2}}{4}\int d^{9}\xi\sqrt{-\det G}e^{-\Phi}\mathrm{Tr}\left\{ -2D_{a}\varphi^{i}D_{a}\varphi^{i}+\left[\varphi^{i},\varphi^{j}\right]^{2}\right\} \nonumber \\
 & =\tilde{T}_{8}\int d^{4}xdz\sqrt{-\det G}e^{-\Phi}\mathrm{Tr}\left\{ \frac{1}{2}D_{a}\Psi D_{a}\Psi-\frac{1}{4}\left[\Psi,\Psi\right]^{2}\right\} .\label{eq:31}
\end{align}
with $D_{a}\Psi=\partial_{a}\Psi+i\left[\mathcal{A}_{a},\Psi\right]$.
According to \cite{key-73}, one can define the moduli by the extrema
of the potential contribution or $\left[\Psi,\left[\Psi,\Psi\right]\right]=0$
in (\ref{eq:31}). So the moduli solution of $\Psi$ for $N_{f}$
light branes separated from one heavy brane can be defined with a
finite vev $v$ as,

\begin{equation}
\Psi=\left(\begin{array}{cc}
-\frac{v}{N_{f}}\boldsymbol{1}_{N_{f}} & 0\\
0 & v
\end{array}\right),\label{eq:32}
\end{equation}
 With solution (\ref{eq:32}), we have 

\begin{equation}
S_{\Psi}=-\tilde{T}U_{KK}^{-1}v^{2}\frac{2\left(N_{f}+1\right)^{2}}{N_{f}^{2}}\int d^{4}xdzH_{0}^{3/2}U^{2}\left(g^{zz}\varPhi_{z}^{\dagger}\varPhi_{z}+g^{\mu\nu}\varPhi_{\mu}^{\dagger}\varPhi_{\nu}\right).\label{eq:33}
\end{equation}
Again, we introduce the dimensionless variable $\varPhi_{a}$ by imposing
$z\rightarrow zU_{KK}$, $x^{\mu}\rightarrow x^{\mu}/M_{KK}$, $\boldsymbol{\mathcal{A}}_{z}\rightarrow\boldsymbol{\mathcal{A}}_{z}/U_{KK}$,
$\boldsymbol{\mathcal{A}}_{\mu}\rightarrow\boldsymbol{\mathcal{A}}_{\mu}M_{KK}$
which means it requires an additional replacement $v\rightarrow\frac{M_{KK}^{1/2}}{U_{KK}^{1/2}}v$.
Then using the $\lambda$ rescaling as in (\ref{eq:27}), we finally
obtain,

\begin{equation}
S_{\Psi}=-aN_{c}b^{3/2}\int d^{4}xdz2m_{H}^{2}\varPhi_{M}^{\dagger}\varPhi_{M}+\mathcal{O}\left(\lambda^{-1}\right),\label{eq:34}
\end{equation}
where $m_{H}=\frac{1}{\sqrt{6}}\frac{N_{f}+1}{N_{f}}vb^{1/4}$. 

For a Dp-brane, there is a CS term in the total action whose standard
form is,

\begin{equation}
S_{\mathrm{CS}}^{D_{\mathrm{p}}}=\mu_{p}\int_{D_{p}}\sum_{q}C_{q+1}\wedge\mathrm{Tr}e^{2\pi\alpha^{\prime}\mathcal{F}}=\mu_{p}\int_{D_{p}}\sum_{n}C_{p-2n+1}\wedge\frac{1}{n!}\left(2\pi\alpha^{\prime}\right)^{n}\mathrm{Tr}\mathcal{F}^{n}.
\end{equation}
In our D0-D4 background, the non-vanished terms for the probe $\mathrm{D}8/\overline{\mathrm{D}8}$-branes
are,

\begin{equation}
S_{\mathrm{CS}}^{\mathrm{D}8/\overline{\mathrm{D}8}}=\frac{1}{3!}\left(2\pi\alpha^{\prime}\right)^{3}\mu_{8}\int_{\mathrm{D}8/\overline{\mathrm{D}8}}C_{3}\wedge\mathrm{Tr}\left[\mathcal{F}^{3}\right]+2\pi\alpha^{\prime}\mu_{8}\int_{\mathrm{D}8/\overline{\mathrm{D}8}}C_{7}\wedge\mathrm{Tr}\left[\mathcal{F}\right].\label{eq:36}
\end{equation}
The first term in (\ref{eq:36}) can be integrated out by using $dC_{3}=f_{4}$
which has been given in (\ref{eq:2}), it yields a CS 5-form,

\begin{equation}
S_{\mathrm{CS}}^{\mathrm{D}8/\overline{\mathrm{D}8}}=\frac{N_{c}}{24\pi^{2}}\int_{\mathbb{R}^{4+1}}\boldsymbol{\mathcal{A}}\boldsymbol{\mathcal{F}}^{2}-\frac{1}{2}\boldsymbol{\mathcal{A}}^{3}\boldsymbol{\mathcal{F}}+\frac{1}{10}\boldsymbol{\mathcal{A}}^{5},\label{eq:37}
\end{equation}
and this term is invariant under the $\lambda$ rescaling (\ref{eq:27}).
However explicit calculations show that the second term in (\ref{eq:36})
becomes $\mathcal{O}\left(\lambda^{-1}\right)$ in the large $\lambda$
limit. So on the light flavor branes, only (\ref{eq:37}) in the CS
term survives in strongly coupled limit. Inserting (\ref{eq:21})
and (\ref{eq:22}) into (\ref{eq:37}) with the dimensionless variables,
(\ref{eq:37}) becomes, 

\begin{equation}
\mathcal{L}_{\mathrm{CS}}^{\mathrm{D}8/\overline{\mathrm{D}8}}=\mathcal{L}_{\mathrm{CS}}^{L}\left(\mathcal{A}\right)+\mathcal{L}_{\mathrm{CS}}^{H},
\end{equation}
where $\mathcal{L}_{\mathrm{CS}}^{L}\left(\mathcal{A}\right)$ represents
the CS term for the light hadrons given in (\ref{eq:77}) (\ref{eq:78})
which has been studied in \cite{key-44,key-50} and

\begin{align}
\mathcal{L}_{\mathrm{CS}}^{H}= & -\frac{iN_{c}}{24\pi^{2}}\left(d\varPhi^{\dagger}\mathcal{A}d\varPhi+d\varPhi^{\dagger}d\mathcal{A}\varPhi+\varPhi^{\dagger}d\mathcal{A}d\varPhi\right)\nonumber \\
 & -\frac{iN_{c}}{16\pi^{2}}\left(d\varPhi^{\dagger}\mathcal{A}^{2}\varPhi+\varPhi^{\dagger}\mathcal{A}^{2}d\varPhi+\varPhi^{\dagger}\mathcal{A}d\mathcal{A}\varPhi+\varPhi^{\dagger}d\mathcal{A}\mathcal{A}\varPhi\right)\nonumber \\
 & -\frac{5iN_{c}}{48\pi^{2}}\varPhi^{\dagger}\mathcal{A}^{3}\varPhi+\mathcal{O}\left(\varPhi^{4},\mathcal{A}\right).\label{eq:39}
\end{align}
Therefore the action for light-heavy interaction can be collected
from (\ref{eq:29}) (\ref{eq:30}) (\ref{eq:34}) (\ref{eq:39}) on
the light flavor branes.

\section{The zero modes}

In the limit of $\lambda\rightarrow\infty$ followed by $m_{H}\rightarrow\infty$,
the heavy meson in bulk can be treated as the instanton configuration
on the flavor branes which can be effectively treated as a spinor.
And it forms a 4-dimensionally flavored zero-mode which can be interpreted
as either a bound of heavy flavor or anti-heavy flavor in the spacetime
of $\left\{ x^{\mu}\right\} $. However, in the Skyrme model, the
Wess-Zumino-Witten term is time-odd which carries opposite signs for
heavy particles and anti- particles. While this is difficult for anti-particles
by holography, it is remarkable.

\subsection{Equations of motion}

Let us consider the solution of the heavy meson field $\varPhi_{M}$.
Notice that $\varPhi_{M}$ is independent of $\varPhi_{0}$, so the
equations of motion read from the action (\ref{eq:29}) (\ref{eq:30})
(\ref{eq:34}) (\ref{eq:39}),

\begin{equation}
D_{M}D_{M}\varPhi_{N}-D_{N}D_{M}\varPhi_{M}+2\mathcal{F}_{NM}\varPhi_{M}+\mathcal{O}\left(\lambda^{-1}\right)=0.\label{eq:40}
\end{equation}
And the equation of motion for $\varPhi_{0}$ is,

\begin{equation}
D_{M}\left(D_{0}\varPhi_{M}-D_{M}\varPhi_{0}\right)-\mathcal{F}^{0M}\varPhi_{M}-\frac{1}{64\pi^{2}ab^{3/2}}\epsilon_{MNPQ}\mathcal{K}_{MNPQ}+\mathcal{O}\left(\lambda^{-1}\right)=0,\label{eq:41}
\end{equation}
where the 4-form $\mathcal{K}_{MNPQ}$ is given as,

\begin{equation}
\mathcal{K}_{MNPQ}=\partial_{M}\mathcal{A}_{N}\partial_{P}\varPhi_{Q}+\mathcal{A}_{M}\mathcal{A}_{N}\partial_{P}\varPhi_{Q}+\partial_{M}\mathcal{A}_{N}\mathcal{A}_{P}\varPhi_{Q}+\frac{5}{6}\mathcal{A}_{M}\mathcal{A}_{N}\mathcal{A}_{P}\varPhi_{Q}.
\end{equation}
In the heavy quark limit, we follow \cite{key-61} to redefine $\varPhi_{M}=\phi_{M}e^{-im_{H}x^{0}}$
for particles while it follows the replacement $m_{H}\rightarrow-m_{H}$
for the anti-particle.

\subsection{The double limit}

It is very difficult to calculate all the contributions from the heavy
meson field $\varPhi_{M}$. Hence we consider the limit of $\lambda\rightarrow\infty$
followed by $m_{H}\rightarrow\infty$ (i.e. named as the ``double
limit''). So the leading contributions come from the light effective
action presented in (\ref{eq:5}) which is of order $\lambda m_{H}^{0}$,
while the next leading contributions come from the heavy-light interaction
Lagrangian $\mathcal{L}_{1}^{H}$ in (\ref{eq:29}) and $\mathcal{L}_{\mathrm{CS}}^{H}$
in (\ref{eq:39}) which is of order $\lambda^{0}m_{H}$. The double
limit is valid if we assume the heavy meson field $\varPhi_{M}$ is
very massive which means the separation of the heavy and light branes
is very large as shown in Figure \ref{fig:1}. So the straight pending
string takes a value at $z=z_{H}$ which satisfies,

\begin{align}
m_{H} & =\frac{1}{\pi l_{s}^{2}}\lim_{z_{H}\rightarrow\infty}\int_{0}^{z_{H}}dz\sqrt{-g_{00}g_{zz}}\nonumber \\
 & \simeq\frac{1}{\pi l_{s}^{2}}U_{KK}^{1/3}z_{H}^{2/3}+\mathcal{O}\left(z_{H}^{0}\right).\label{eq:43}
\end{align}
It would be convenient to rewrite (\ref{eq:43}) with the dimensionless
variables by the replacement $m_{H}\rightarrow m_{H}M_{KK}$, $z_{H}\rightarrow z_{H}U_{KK}$,
then using (\ref{eq:13}) we have,

\begin{equation}
\frac{m_{H}}{\lambda}=\frac{2b}{9\pi}z_{H}^{2/3}.\label{eq:44}
\end{equation}
According to the above discussion, the derivative of $\varPhi_{M}$
can be replaced by $D_{0}\varPhi_{M}\rightarrow\left(D_{0}\pm im_{H}\right)\varPhi_{M}$
with ``-'' for particle and ``+'' for anti-particle. Then we collect
the order $\lambda^{0}m_{H}$ from our heavy-light action which is,

\begin{align}
\mathcal{L}_{m_{H}}= & \mathcal{L}_{1,m}+\mathcal{L}_{\mathrm{CS},m},\nonumber \\
\mathcal{L}_{1,m}= & ab^{3/2}N_{c}\left[4im_{H}\phi_{M}^{\dagger}D_{0}\phi_{M}-2im_{H}\left(\phi_{0}^{\dagger}D_{M}\phi_{M}-c.c.\right)\right],\nonumber \\
\mathcal{L}_{\mathrm{CS},m}= & \frac{m_{H}N_{c}}{16\pi^{2}}\epsilon_{MNPQ}\phi_{M}^{\dagger}\mathcal{F}_{NP}\phi_{Q}=\frac{m_{H}N_{c}}{8\pi^{2}}\phi_{M}^{\dagger}\mathcal{F}_{MN}\phi_{N}.\label{eq:45}
\end{align}
The equation of motion (\ref{eq:41}) suggests a considerable simplification
$D_{M}\varPhi_{M}=0$ which implies that $\varPhi_{M}$ is covariantly
transverse mode.

\subsection{Vector to spinor}

In the $N_{f}=2$ case of the D0-D4/D8 system, the small size instanton
is described by a flat-space 4-dimensional instanton solution of $SU\left(2\right)$
Yang-mills theory \cite{key-50} in the large $\lambda$ limit, which
is,

\begin{align}
\mathcal{A}_{M}^{cl}= & -\bar{\sigma}_{MN}\frac{x^{N}}{x^{2}+\rho^{2}},\nonumber \\
\mathcal{A}_{0}^{cl}= & -\frac{i}{8\pi^{2}ab^{3/2}x^{2}}\left[1-\frac{\rho^{4}}{\left(x^{2}+\rho^{2}\right)^{2}}\right],\label{eq:46}
\end{align}
where $x^{2}=\left(x^{M}-X^{M}\right)^{2}$ and $X^{M}$ is the constant.
Notice that in (\ref{eq:46}), $\mathcal{A}_{0}^{cl}$ is Abelian
while $\mathcal{A}_{M}^{cl}$ is non-Abelian. It carries a field strength,

\begin{equation}
\mathcal{F}_{MN}=\frac{2\bar{\sigma}_{MN}\rho^{2}}{\left(x^{2}+\rho^{2}\right)^{2}}.\label{eq:47}
\end{equation}
By defining $f_{MN}=\partial_{[M}\phi_{N]}+\mathcal{A}_{[M}\phi_{N]}$,
$\mathcal{L}_{0}^{H}$ in (\ref{eq:29}) can be rewritten as follows,

\begin{align}
\mathcal{L}_{0}^{H} & =-f_{MN}^{\dagger}f_{MN}+2\phi_{M}^{\dagger}\mathcal{F}_{MN}\phi_{N}\nonumber \\
 & =-f_{MN}^{\dagger}f_{MN}+2\epsilon_{MNPQ}\phi_{M}^{\dagger}D_{N}D_{P}\phi_{Q}\nonumber \\
 & =-f_{MN}^{\dagger}f_{MN}+f_{MN}^{\dagger}\star f_{MN}\nonumber \\
 & =-\frac{1}{2}\left(f_{MN}-\star f_{MN}\right)^{\dagger}\left(f_{MN}-\star f_{MN}\right),\label{eq:48}
\end{align}
where $\star$ represents the Hodge dual. Therefore the equations
of motion (\ref{eq:40}) can be replaced by,

\begin{align}
f_{MN}-\star f_{MN}=0 & ,\nonumber \\
D_{M}\phi_{M}=0 & ,\label{eq:49}
\end{align}
which is equivalent to

\begin{equation}
\sigma_{M}D_{M}\psi=0,\ \ \ \ \mathrm{with}\ \ \psi=\bar{\sigma}_{M}\phi_{M}.\label{eq:50}
\end{equation}
So $\phi_{M}$ can be solved from (\ref{eq:49}) as,

\begin{equation}
\phi_{M}=\bar{\sigma}_{M}\xi\frac{\rho}{\left(x^{2}+\rho^{2}\right)^{3/2}}\equiv\bar{\sigma}_{M}f\left(x\right)\xi,
\end{equation}
which is in agreement with \cite{key-61}. $\xi$ is a two-component
spinor and the interplay of (\ref{eq:50}) is remarkable since it
shows that a heavy vector meson holographically binds to an instantonic
configuration in bulk which concludes that a vector zero mode is equivalently
described by a spinor.

\section{Quantization}

The classical moduli of the bound instanton zero-mode should be quantized
by slowly rotating and translating the bound state since it breaks
rotational and translational symmetry. The quantization of the leading
$\lambda N_{c}$ contribution can be found in \cite{key-50} which
is instantonic and standard, while the quantization of the sub-leading
$\lambda^{0}m_{H}$ contribution involving zero-modes in the D0-D4/D8
system is new. We will employ the quantization applied on D4/D8 as
\cite{key-61}.

\subsection{Collectivization}

As in \cite{key-5,key-9}, we assume that the zero-modes slowly rotates,
translates and deforms through

\begin{align}
\varPhi_{M} & \rightarrow V\left[a_{I}\left(t\right)\right]\varPhi_{M}\left[X^{0}\left(t\right),Z\left(t\right),\rho\left(t\right),\chi\left(t\right)\right],\nonumber \\
\varPhi_{0} & \rightarrow0+\delta\phi_{0},\label{eq:52}
\end{align}
where $X^{0}$, $Z$ is the center in the $x^{i}$ and $z$ directions
respectively. $a_{I}$ is the $SU\left(2\right)$ gauge rotation.
They are represented by $X^{\alpha}=\left(X^{i},Z,\rho\right)$ with

\begin{align}
-iV^{\dagger}\partial_{0}V & =\Phi=-\partial_{t}X^{M}\mathcal{A}_{M}+\chi^{i}\varPhi_{i},\nonumber \\
\chi^{i} & =-i\mathrm{Tr}\left(\tau^{i}a_{I}^{-1}\partial_{t}a_{I}\right).\label{eq:53}
\end{align}
Here $a^{I}$s carry the quantum numbers of isospin, the angular momentum
and $\tau^{i}$s are Pauli matrices. Since the equation (\ref{eq:41})
has to be satisfied, $\delta\phi_{0}$ is fixed at the next-leading
order,

\begin{align}
 & -D_{M}^{2}\delta\phi_{0}+D_{M}\bar{\sigma}_{M}\left[\partial_{t}X^{i}\frac{\partial\left(f\chi\right)}{\partial X^{i}}+\partial_{t}\chi\right]\nonumber \\
 & +i\left(\partial_{t}X^{\alpha}\partial_{\alpha}\varPhi_{M}-D_{M}\Phi\right)\bar{\sigma}_{M}\chi+\delta S_{\mathrm{CS}}=0.\label{eq:54}
\end{align}
For a general quantization of the ensuing moduli, we can solve (\ref{eq:54})
and then insert the solution back into the action.

\subsection{Leading order of the heavy mass term}

The heavy mass terms in the double limit are given in (\ref{eq:45}).
Imposing (\ref{eq:52}) (\ref{eq:53}) on (\ref{eq:45}), the contributions
to order $\lambda^{0}m_{H}$ in $\mathcal{L}_{1,m}$ come from three
terms which are,

\begin{equation}
\mathcal{L}_{1,m}=ab^{3/2}N_{c}\left(16im_{H}\xi^{\dagger}\partial_{t}\xi f^{2}+16im_{H}\xi^{\dagger}\xi\mathcal{A}_{0}f^{2}-16m_{H}f^{2}\xi^{\dagger}\sigma_{\mu}\Phi\bar{\sigma}_{\mu}\xi\right),\label{eq:55}
\end{equation}
where $\mathcal{A}_{0}$ is the rescaled $U\left(1\right)$ gauge
field. With the gauge field strength (\ref{eq:47}), the CS term in
(\ref{eq:45}) can be written as,

\begin{equation}
\mathcal{L}_{\mathrm{CS},m}=\frac{3m_{H}N_{c}}{\pi^{2}}\frac{f^{2}\rho^{2}}{\left(x^{2}+\rho^{2}\right)^{2}}\xi^{\dagger}\xi.\label{eq:56}
\end{equation}
Notice that the third term in (\ref{eq:55}) vanishes owing to the
identity $\sigma_{\mu}\tau^{i}\bar{\sigma}_{\mu}=0$. 

There is a Coulomb-like backreaction according to the coupling $\xi^{\dagger}\xi\mathcal{A}_{0}$
in (\ref{eq:55}). To clarify this, let us introduce a Coulomb-like
potential defined as $\varphi=-i\mathcal{A}_{0}$. We collect all
the $U\left(1\right)$ coupling from (\ref{eq:45}) up to $\mathcal{O}\left(\lambda^{0}m_{H}\right)$
as,

\begin{equation}
\mathcal{L}_{U\left(1\right)}=-ab^{3/2}N_{c}\left[\frac{1}{2}\left(\nabla\varphi\right)^{2}+\varphi\left(\rho_{0}-16m_{H}f^{2}\xi^{\dagger}\xi\right)\right],\label{eq:57}
\end{equation}
where $\rho_{0}$ is the $U\left(1\right)$ ``charge'' which is
given as,

\begin{equation}
\rho_{0}=\frac{1}{64\pi^{2}ab^{3/2}}\epsilon_{MNPQ}\mathcal{F}_{MN}\mathcal{F}_{PQ}.
\end{equation}
Solving the equation of motion from (\ref{eq:57}) for $\varphi$,
one obtains its onshell action as,

\begin{equation}
\mathcal{L}_{U\left(1\right)}=\mathcal{L}_{U\left(1\right)}\left[\rho_{0}\right]+16ab^{3/2}N_{c}m_{H}f^{2}\xi^{\dagger}\xi\left(-i\mathcal{A}_{0}^{cl}\right)-\frac{ab^{3/2}N_{c}}{24\pi^{2}\rho^{2}}\left(16m_{H}\xi^{\dagger}\xi\right)^{2}.\label{eq:59}
\end{equation}
The last term is the Coulomb-like self-interaction which is repulsive
and tantamount of fermion number repulsion in holography.

\subsection{Moduli effective action}

All the contributions up to $\mathcal{O}\left(\lambda^{0}m_{H}\right)$
in the effective moduli action can be collected from (\ref{eq:55})
(\ref{eq:57}) and (\ref{eq:59}). Let us summarize them as follows,

\begin{align}
\mathcal{L}= & \mathcal{L}^{L}\left[a_{I},X^{\alpha}\right]+16im_{H}ab^{3/2}N_{c}\xi^{\dagger}\partial_{t}\xi\int d^{4}xf^{2}\nonumber \\
 & -16m_{H}ab^{3/2}N_{c}\xi^{\dagger}\xi\int d^{4}x\left[i\mathcal{A}_{0}^{cl}f^{2}-\frac{3}{16\pi^{2}ab^{3/2}}\frac{f^{2}\rho^{2}}{\left(x^{2}+\rho^{2}\right)^{2}}\right]\nonumber \\
 & -\frac{ab^{3/2}N_{c}}{24\pi^{2}\rho^{2}}\left(16m_{H}\xi^{\dagger}\xi\right)^{2}.\label{eq:60}
\end{align}
Here $\mathcal{L}^{L}$ refers to the effective action on the moduli
space from the contribution of the light hadrons which is identical
to the derivation in \cite{key-50}. In (\ref{eq:60}), it shows the
explicitly new contribution due to the bound heavy meson. In the leading
order, the coupling of the light collective degrees of freedom should
be a general reflection on heavy quark symmetry. However there is
no such coupling in the order of $\mathcal{O}\left(\lambda^{0}m_{H}\right)$
in (\ref{eq:60}) to the heavy spinor degree of freedom $\xi$. Notice
that the coupling to the instanton size $\rho$ does not upset this
symmetry. In order to calculate (\ref{eq:60}), we follow the steps
as \cite{key-12} i.e. using the normalization $\int d^{4}xf^{2}=1$,
inserting the explicit form of $\mathcal{A}_{0}^{cl}$, and rescaling
$\xi\rightarrow\xi/\sqrt{16ab^{3/2}N_{c}m_{H}}$. Finally it yields,

\begin{equation}
\mathcal{L}=\mathcal{L}^{L}\left[a_{I},X^{\alpha}\right]+i\xi^{\dagger}\partial_{t}\xi+\frac{3}{32\pi^{2}ab^{3/2}\rho^{2}}\xi^{\dagger}\xi-\frac{\left(\xi^{\dagger}\xi\right)^{2}}{24\pi^{2}ab^{3/2}\rho^{2}N_{c}}.\label{eq:61}
\end{equation}
It shows the zero-mode of the vector to the instanton transmutation
of a massive spinor with a repulsively Coulomb-like self-interaction
in the presence of the D0 charge. A negative mass term also means
the heavy meson lowers its energy. So the preceding arguments are
also suitable for an anti-heavy meson in the presence of an instanton
with a positive mass term leading to (\ref{eq:61}). The energy of
this meson rises in the presence of the instanton to order $\lambda^{0}$.
It originates from the Chern-Simons term in the holographical action
which is the analogue of the effects due to the Wess-Zumino-Witten
term in the Skyrme model.

\subsection{Heavy-light spectrum}

The step to quantize the Lagrangian (\ref{eq:60}) follows the same
discussion as those presented in \cite{key-50} for $\mathcal{L}^{L}\left[a_{I},X^{\alpha}\right]$.
We use $\mathcal{H}^{L}\left[a_{I},X^{\alpha}\right]$ to represent
the Hamiltonian associated to $\mathcal{L}^{L}\left[a_{I},X^{\alpha}\right]$,
then the Hamiltonian for (\ref{eq:61}) takes the following formula,

\begin{equation}
\mathcal{H}=\mathcal{H}^{L}\left[a_{I},X^{\alpha}\right]-\frac{3}{32\pi^{2}ab^{3/2}\rho^{2}}\xi^{\dagger}\xi+\frac{\left(\xi^{\dagger}\xi\right)^{2}}{24\pi^{2}ab^{3/2}\rho^{2}N_{c}}.\label{eq:62}
\end{equation}
And the quantization rule for the spinor $\xi$ should be chosen as,

\begin{equation}
\xi_{i}\xi_{j}^{\dagger}+\xi_{j}^{\dagger}\xi_{i}=\delta_{ij}.
\end{equation}
So the rotation of the spinor $\xi$ is equivalent to a spatial rotation
of the heavy vector meson field $\phi_{M}$ since $U^{-1}\bar{\sigma}_{M}U=\Lambda_{MN}\bar{\sigma}_{M}$,
where $U$ and $\Lambda$ represents the rotation of a spinor and
a vector respectively e.g. $\xi\rightarrow U\xi,\ \phi_{M}\rightarrow\Lambda_{MN}\phi_{N}$.
The parity of $\xi$ is positive which is opposite to $\phi_{M}$. 

The spectrum of (\ref{eq:62}) follows the same discussion in \cite{key-50}.
Since the (\ref{eq:62}) contains only two terms proportional to $\rho^{-2}$,
by comparing (\ref{eq:62}) with the $\mathcal{H}^{L}\left[a_{I},X^{\alpha}\right]$
presented in \cite{key-50}, the heavy-light spectrum can therefore
be obtained by modifying $Q$ as,

\begin{equation}
Q=\frac{N_{c}}{40ab^{3/2}\pi^{2}}\rightarrow\frac{N_{c}}{40ab^{3/2}\pi^{2}}\left[1-\frac{15}{4N_{c}}\xi^{\dagger}\xi+\frac{5}{3N_{c}^{2}}\left(\xi^{\dagger}\xi\right)^{2}\right].
\end{equation}
Let us use $\boldsymbol{\mathrm{J}}$ and $\boldsymbol{\mathrm{I}}$
to represent the spin and isospin, so they are related by

\begin{equation}
\vec{\boldsymbol{\mathrm{J}}}=-\vec{\boldsymbol{\mathrm{I}}}+\vec{\boldsymbol{\mathrm{S}}}=-\vec{\boldsymbol{\mathrm{I}}}+\xi^{\dagger}\frac{\vec{\tau}}{2}\xi.
\end{equation}
And notice that we have $\boldsymbol{\mathrm{J}}+\boldsymbol{\mathrm{I}}=0$
in the absence of the heavy-light meson as expected from the spin-flavor
hedgehog character. The quantum states for a single bound state i.e.
$N_{Q}\equiv\xi^{\dagger}\xi=1$ and $IJ^{\pi}$ assignments are labeled
by,

\begin{equation}
|N_{Q},J_{M},l_{m},n_{z},n_{\rho}>\ \ with\ \ IJ^{\pi}=\frac{l}{2}\left(\frac{l}{2}\pm\frac{1}{2}\right)^{\pi}.
\end{equation}
Here $n_{z},n_{\rho}=0,1,2...$ represents the number of quanta associated
to the collective motion and the radial breathing of the instanton
core respectively. Following \cite{key-44,key-50}, the spectrum of
the bound heavy-light state in D0-D4/D8 system is,

\begin{align}
M_{N_{Q}}= & M_{0}+N_{Q}m_{H}+M_{KK}\sqrt{\frac{3-b}{3}}\left(n_{\rho}+n_{z}+1\right)\nonumber \\
 & +M_{KK}\left[\frac{\left(l+1\right)^{2}\left(3-b\right)}{12}+\frac{3-b}{15}N_{c}^{2}\left(1-\frac{15}{4N_{c}}N_{Q}+\frac{5}{3N_{c}^{2}}N_{Q}^{2}\right)\right]^{1/2}.
\end{align}
$M_{KK}$ is the Kaluza-Klein mass and $M_{0}=\frac{\lambda N_{c}b^{3/2}}{27\pi}M_{KK}$.

\subsubsection*{Single heavy-baryon spectrum}

The lowest heavy states with one heavy quark are characterized by
$N_{Q}=1,l=even,N_{c}=3$ and $n_{z},n_{\rho}=0,1$. So the mass spectrum
is given as,

\begin{align}
M_{single}= & M_{0}+m_{H}+M_{KK}\left(n_{\rho}+n_{z}+1\right)\sqrt{\frac{3-b}{3}}\nonumber \\
 & +M_{KK}\left[\frac{\left(l+1\right)^{2}\left(3-b\right)}{12}-\frac{7}{180}\left(3-b\right)\right]^{1/2}.\label{eq:68}
\end{align}
Let us consider the states with $n_{z}=n_{\rho}=0$ and identify the
state with $l=0$, the assignments $IJ^{\pi}=0\frac{1}{2}^{+}$ as
the heavy-light iso-singlet $\Lambda_{Q}$. Then we identify the state
with $l=2$ and the assignments $IJ^{\pi}=1\frac{1}{2}^{+},1\frac{3}{2}^{+}$
as the heavy-light iso-triplet $\Sigma_{Q},\Sigma_{Q}^{*}$ respectively.
Subtracting the nucleon mass $M_{N}$ (which is identified as the
state with $l=0$ of the light-baryon spectrum) from (\ref{eq:68})
, we have,

\begin{align}
M_{\Lambda_{Q}}-M_{N}-m_{H} & \simeq-0.76\sqrt{3-b}M_{KK},\nonumber \\
M_{\Sigma_{Q}}-M_{N}-m_{H} & \simeq-0.12\sqrt{3-b}M_{KK},\nonumber \\
M_{\Sigma_{Q}^{*}}-M_{N}-m_{H} & \simeq-0.12\sqrt{3-b}M_{KK}.
\end{align}
Thus we see the explicit dependence of D0 charge in the baryon spectrum
in this model. Next we can study the excited heavy baryons with (\ref{eq:68}).
Let us consider the low-lying breathing modes $R$ ($n_{\rho}=1$)
with the even assignments $IJ^{\pi}=0\frac{1}{2}^{+},1\frac{1}{2}^{+},1\frac{3}{2}^{+}$
and the odd parity excited states $O$ ($n_{z}=1$) with the even
assignments $IJ^{\pi}=0\frac{1}{2}^{-},1\frac{1}{2}^{-},1\frac{3}{2}^{-}$.
Using (\ref{eq:68}), we have ($E=O,R$),

\begin{align}
M_{\Lambda_{EQ^{\prime}}}\left(b\right) & =+0.23M_{\Lambda_{Q}}\left(b\right)+0.77M_{N}\left(b\right)-0.23m_{H}+m_{H}^{\prime},\nonumber \\
M_{\Sigma_{EQ^{\prime}}}\left(b\right) & =-0.59M_{\Lambda_{Q}}\left(b\right)+1.59M_{N}\left(b\right)+0.59m_{H}+m_{H}^{\prime},\label{eq:70}
\end{align}
where the holographically model-independent relations in \cite{key-61},

\begin{align}
M_{\Lambda_{Q^{\prime}}} & =M_{\Lambda_{Q}}+m_{H^{\prime}}-m_{H},\nonumber \\
M_{\Sigma_{Q^{\prime}}} & =0.84M_{N}+m_{H^{\prime}}+0.16\left(M_{\Lambda_{Q}}-m_{H}\right),
\end{align}
has been imposed.

\subsubsection*{Double-heavy baryons}

Since heavy baryons also contain anti-heavy quarks, let us return
to the preceding arguments using the reduction $\varPhi_{M}=\phi_{M}e^{+im_{H}x^{0}}$
, in order to amount an anti-heavy-light meson. Most of the calculations
are similar except for pertinent minus signs to the effective Lagrangian.
In the form of a zero-mode, if we bind one heavy-light and one anti-heavy-light
meson, the effective Lagrangian now reads,

\begin{align}
\mathcal{L}= & \mathcal{L}^{L}\left[a_{I},X^{\alpha}\right]+i\xi_{Q}^{\dagger}\partial_{t}\xi_{Q}+\frac{3}{32\pi^{2}ab^{3/2}\rho^{2}}\xi_{Q}^{\dagger}\xi_{Q}\nonumber \\
 & -i\xi_{\bar{Q}}^{\dagger}\partial_{t}\xi_{\bar{Q}}-\frac{3}{32\pi^{2}ab^{3/2}\rho^{2}}\xi_{\bar{Q}}^{\dagger}\xi_{\bar{Q}}+\frac{\left(\xi_{Q}^{\dagger}\xi_{Q}-\xi_{\bar{Q}}^{\dagger}\xi_{\bar{Q}}\right)^{2}}{24\pi^{2}ab^{3/2}\rho^{2}N_{c}}.
\end{align}
The contributions of the mass from a heavy-light and anti-heavy-light
meson are opposite as we have indicated. So the mass spectrum for
baryons with $N_{Q}$ heavy- quarks and $N_{\bar{Q}}$ anti-heavy
quarks can be calculated as,

\begin{align}
M_{Q\bar{Q}}= & M_{0}+\left(N_{Q}+N_{\bar{Q}}\right)m_{H}+M_{KK}\sqrt{\frac{3-b}{3}}\left(n_{\rho}+n_{z}+1\right)\nonumber \\
 & +M_{KK}\left\{ \frac{\left(l+1\right)^{2}\left(3-b\right)}{12}+\frac{3-b}{15}N_{c}^{2}\left[1-\frac{15\left(N_{Q}-N_{\bar{Q}}\right)}{4N_{c}}+\frac{5\left(N_{Q}-N_{\bar{Q}}\right)^{2}}{3N_{c}^{2}}\right]\right\} ^{1/2}.
\end{align}
The lowest state ($N_{Q}=N_{\bar{Q}}=1,n_{\rho}=n_{z}=0,l=1$) with
the assignments $IJ^{\pi}=\frac{1}{2}\frac{1}{2}^{-},\frac{1}{2}\frac{3}{2}^{-}$
can be identified as pentaquark baryonic states and the masses are
given as,

\begin{equation}
M_{Q\bar{Q}}\left(b\right)-M_{N}\left(b\right)-2m_{H}=0
\end{equation}
which does not depend on the D0 charge obviously. 

For the excited pentaquark states, we identify the lowest state as
$O$ with the odd parity, assignments $IJ^{\pi}=\frac{1}{2}\frac{1}{2}^{+},\frac{1}{2}\frac{3}{2}^{+}$
and quantum number $N_{Q}=N_{\bar{Q}}=1,n_{\rho}=0,n_{z}=1,l=1$.
The state with quantum number $N_{Q}=N_{\bar{Q}}=1,n_{\rho}=1,n_{z}=0,l=1$
and same assignments is identified as breathing or Roper $R$ pentaquarks
as the ground state. So the mass relations for these states are given
as ($E=O,R$),

\begin{equation}
M_{EQ\bar{Q}}\left(b\right)-M_{N}\left(b\right)-2m_{H}\simeq0.58\sqrt{3-b}M_{KK}.
\end{equation}
On the other hand, the Delta-type pentaquarks can be identified as
the states with quantum number $N_{Q}=N_{\bar{Q}}=1,n_{\rho}=n_{z}=0,l=3$.
Altogether, we have one $IJ^{\pi}=\frac{3}{2}\frac{1}{2}^{-}$, two
$IJ^{\pi}=\frac{3}{2}\frac{3}{2}^{-}$ , and one $IJ^{\pi}=\frac{3}{2}\frac{5}{2}^{-}$
states, so the masses with heavy-flavors are given as,

\begin{equation}
M_{\Delta Q\bar{Q}}\left(b\right)-M_{N}\left(b\right)-2m_{H}\simeq0.42\sqrt{3-b}M_{KK}.
\end{equation}

\section{Summary}

Using the Witten-Sakai-Sugimoto model in the D0-D4 background \cite{key-48,key-49}
and the mechanism proposed in \cite{key-59,key-60,key-61},  we have
extended the analysis in \cite{key-50,key-51} to involve the heavy
flavors by a top-down holographic approach to the single- and double-heavy
baryon spectra. The heavy-light interaction is introduced into this
model by considering a pair of heavy flavor brane which is separated
from the light flavor branes. The heavy baryon emerges from the zero
mode of the reduced vector meson field to order $\lambda m_{H}^{0}$.
The bind of the heavy and anti-heavy meson is equivalent as the instanton
configurations of the gauge field on the flavor branes in leading
order of $\lambda$, even in the presence of the Chern-Simons term.
The smeared D0 charge has been turned on in the D4-soliton background,
so our calculation contains the excited states with nonzero $\mathrm{Tr}\left[\mathcal{F}\wedge\mathcal{F}\right]$
or a nonzero $\theta$ angle in the dual field theory. The $\theta$
dependence is through a parameter $b$ (or $\tilde{\kappa}$) which
is monotonically increasing with $\theta$.

Following the quantization in \cite{key-61}, the bound state moduli
gives a rich spectrum. It contains the coupled rotational, vibrational
and translational modes. There are also some newly excited states
in the spectrum which are yet to be observed. The charmed pentaquark
can be naturally identified as a pair of degenerate heavy iso-doublets
with $IJ^{\pi}=\frac{1}{2}\frac{1}{2}^{-},\frac{1}{2}\frac{3}{2}^{-}$
in the spectra when it is extended to the double-heavy baryon case.
Our calculation also shows the D0 charge moduli in some new pentaquarks
with hidden charm and bottom, and five Delta-like pentaquarks with
hidden charm in the spectra. Notice that our discussion returns to
those in \cite{key-61} if $b=1$ i.e. no D0 charge. Particularly
for $b>3$ case, we notice that the spectrum becomes complex which
indicates that baryons cannot be stable and it is in agreement with
the previous study \cite{key-50,key-51} of the holographic baryons
in this model.

As in the most approaches of the gauge-gravity duality, our analysis
is done in the large $N_{c}$, large Hooft coupling $\lambda$ limit,
and now with large $m_{H}$. Since the baryon spectrum demonstrates
behavior of light baryons in \cite{key-50,key-51}, we expect this
model also captures the qualitative $\tilde{\kappa}$ (or $\theta$
angle) behavior, at least, for small $\tilde{\kappa}$ (or $\theta$
angle) in QCD-like theory when the heavy-light interaction is involved.
Although we have compared our results with the real-world nuclei or
quark states by setting $N_{c}=3$ as \cite{key-50,key-51}, there
is still a long way to the realistic baryon spectrum.

\section*{Appendix}

In this Appendix, we collect the essential steps to quantize the light-baryon
Lagrangian $\mathcal{L}^{L}$ which is presented in this manuscript.
The details can be systematically reviewed in \cite{key-44,key-50}.
With the dimensionless variables, the explicit formula of $\mathcal{L}^{L}\left[a_{I},X^{\alpha}\right]$
is given as,\index{Commands!T!tag@\textbackslash{}tag}

\begin{equation}
\mathcal{L}^{L}=\mathcal{L}_{YM}^{L}+\mathcal{L}_{CS}^{L},\tag{A-1}\label{eq:77}
\end{equation}
where

\begin{align}
\mathcal{L}_{YM}^{L} & =-a\lambda N_{c}b^{1/2}\int d^{4}xdzH_{0}^{1/2}\left(U\right)\mathrm{Tr}\left[\frac{1}{2}\frac{U_{KK}}{U}\mathcal{F}_{\mu\nu}\mathcal{F}^{\mu\nu}+\frac{U^{3}}{U_{KK}^{3}}b\mathcal{F}_{\mu z}\mathcal{F}^{\mu z}\right],\nonumber \\
\mathcal{L}_{CS}^{L} & =\frac{N_{c}}{24\pi^{2}}\mathrm{Tr}\left[\mathcal{A}\wedge\mathcal{F}\wedge\mathcal{F}-\frac{1}{2}\mathcal{A}^{3}\wedge\mathcal{F}+\frac{1}{10}\mathcal{A}^{5}\right].\tag{A-2}\label{eq:78}
\end{align}
For the two-flavor case, the $U\left(2\right)$ gauge field $\mathcal{A}$
can be decomposed to a $SU\left(2\right)$ part $A$ and a $U\left(1\right)$
part $\hat{A}$ as,

\begin{equation}
\mathcal{A}=A+\frac{1}{2}\hat{A},\tag{A-3}
\end{equation}
whose gauge field strength is,

\begin{equation}
\mathcal{F}=F+\frac{1}{2}\hat{F}.\tag{A-4}
\end{equation}
In the large $\lambda$ limit, imposing the $\lambda$ rescale (\ref{eq:27}),
the EOM from (\ref{eq:78}) can be obtained as,

\begin{align}
D_{M}F_{MN}+\mathcal{O}\left(\lambda^{-1}\right)=0, & \ \ D_{M}F_{0M}+\frac{1}{64\pi^{2}ab^{3/2}}\epsilon_{MNPQ}\hat{F}_{MN}F_{PQ}+\mathcal{O}\left(\lambda^{-1}\right)=0\nonumber \\
\partial_{M}\hat{F}_{0M}+\mathcal{O}\left(\lambda^{-1}\right)=0, & \ \ \partial_{M}\hat{F}_{0M}+\frac{1}{64\pi^{2}ab^{3/2}}\epsilon_{MNPQ}\mathrm{Tr}\left[F_{MN}F_{PQ}\right]+\mathcal{O}\left(\lambda^{-1}\right)=0,\tag{A-5}
\end{align}
and the solution is given in (\ref{eq:46}). 

In order to obtain the spectrum, we require the moduli of the solution
to be time-dependent, i.e.

\begin{equation}
X^{\alpha},a^{I}\rightarrow X^{\alpha}\left(t\right),a^{I}\left(t\right).\tag{A-6}
\end{equation}
Here $a^{I}\left(t\right)$ refers to the $SU\left(2\right)$ orientation.
So the $SU\left(2\right)$ gauge transformation also becomes time-dependent,

\begin{align}
\mathcal{A}_{M} & \rightarrow V\left(\mathcal{A}_{M}^{cl}-i\partial_{M}\right)V^{-1},\nonumber \\
\mathcal{F}_{MN} & \rightarrow V\mathcal{F}_{MN}^{cl}V^{-1},\ F_{0M}\rightarrow V\left(\dot{X}^{\alpha}\partial_{\alpha}\mathcal{A}_{M}^{cl}-D_{M}^{cl}\Phi\right)V^{-1},\tag{A-7}
\end{align}
where $\Phi=-iV^{\dagger}\partial_{0}V,\ V^{\dagger}=V^{-1}$.

The motion of the collective coordinates could be characterized by
the effective Lagrangian in the moduli space. Up to $\mathcal{O}\left(\lambda^{-1}\right)$,
it is

\begin{align}
L & =\frac{1}{2}m_{X}g_{\alpha\beta}\dot{X}^{\alpha}\dot{X}^{\beta}-U\left(X^{\alpha}\right)+\mathcal{O}\left(\lambda^{-1}\right)\nonumber \\
 & =\frac{1}{2}m_{X}\dot{\vec{X}}^{2}+\frac{1}{2}m_{Z}\dot{Z}^{2}+\frac{1}{2}m_{y}\dot{y}_{I}\dot{y}_{I}-U\left(X^{\alpha}\right),\tag{A-8}\label{eq:84}
\end{align}
where the dot represents the derivative with respect to $t$, $g_{\alpha\beta}$
is the metric of the moduli space parameterized by $X^{\alpha}$ which
satisfies $ds^{2}=g_{\alpha\beta}dX^{\alpha}dX^{\beta}=d\vec{X}^{2}+dZ^{2}+2dy_{I}dy_{I}$
and $\sum_{I=1}^{4}y_{I}y_{I}=\rho^{2}$. $U\left(X^{\alpha}\right)$
is the effective potential associated to the onshell Lagrangian with
the instanton solution (\ref{eq:46}), i.e.

\begin{equation}
\int d^{3}xdz\mathcal{L}^{L}\left[a_{I},X^{\alpha}\right]_{onshell}=-U\left(X^{\alpha}\right).\tag{A-9}
\end{equation}
The baryon spectrum can be obtained by quantizing (\ref{eq:84}) (soliton)
at rest. The quantization procedure is nothing but to replace the
momenta in the Lagrangian to the corresponding differential operators
which can act on the wave function of baryon states. So the quantized
Hamiltonian associated to (\ref{eq:84}) is,

\begin{align}
H & =H_{X}+H_{Z}+H_{y},\nonumber \\
H_{X} & =\frac{1}{2m_{X}}P_{X}^{2}+M_{0}=-\frac{1}{2m_{X}}\sum_{i=1}^{3}\frac{\partial^{2}}{\partial X_{i}^{2}}+M_{0},\nonumber \\
H_{Z} & =\frac{1}{2m_{Z}}P_{Z}^{2}+\frac{1}{2}m_{Z}\omega_{Z}^{2}Z^{2}=-\frac{1}{2m_{Z}}\frac{\partial^{2}}{\partial Z^{2}}+\frac{1}{2}m_{Z}\omega_{Z}^{2}Z^{2},\nonumber \\
H_{y} & =\frac{1}{2m_{y}}P_{y}^{2}+\frac{1}{2}m_{y}\omega_{y}^{2}\rho^{2}+\frac{Q}{\rho^{2}}=-\frac{1}{2m_{y}}\sum_{I=1}^{4}\frac{\partial^{2}}{\partial y_{I}^{2}}+\frac{1}{2}m_{y}\omega_{y}^{2}\rho^{2}+\frac{Q}{\rho^{2}}.\tag{A-10}\label{eq:86}
\end{align}
In the unit of $U_{KK}=M_{KK}=1$ or equivalently, with the replacement
$z\rightarrow zU_{KK}$, $x^{\mu}\rightarrow x^{\mu}/M_{KK}$, $\boldsymbol{\mathcal{A}}_{z}\rightarrow\boldsymbol{\mathcal{A}}_{z}/U_{KK}$,
$\boldsymbol{\mathcal{A}}_{\mu}\rightarrow\boldsymbol{\mathcal{A}}_{\mu}M_{KK}$,
we have the following dimensionless values,

\begin{equation}
M_{0}=8\pi^{2}\lambda ab^{3/2}N_{c},\omega_{Z}=\frac{1}{3}\left(3-b\right),\omega_{y}=\frac{1}{12}\left(3-b\right),Q=\frac{N_{c}}{40\pi^{2}ab^{3/2}},\tag{A-11}
\end{equation}
The eigenstates of $H_{Z}$ are nothing but harmonic-oscillator states.
The eigenfunctions of $H_{y}$ are represented by $T^{l}\left(a_{I}\right)R_{l,n_{\rho}}\left(\rho\right)$
where $T^{l}\left(a_{I}\right)$ are the spherical harmonic functions
on $S^{3}$. They are in the representations of $\left(\frac{l}{2},\frac{l}{2}\right)$
under the transformation of $SO\left(4\right)=SU\left(2\right)\times SU\left(2\right)/Z_{2}$.
The former $SU\left(2\right)$ corresponds to the isometric rotation
while the latter is the space rotation in $\left\{ x^{i}\right\} $.
The states with $I=J=\frac{l}{2}$ are described by this quantization,
so the nucleon state is realized as the lowest state with $l=1,n_{\rho}=n_{z}=0$
of the Hamiltonian (\ref{eq:86}).

\end{document}